\documentclass[aps,prb,twocolumn,showpacs]{revtex4}

\usepackage{graphicx}

\begin{document}

\title{Magnetoresistance of a semiconducting magnetic wire with domain wall}

\author{V. K. Dugaev$^{1,2,}$\cite{email}, J. Barna\'s$^{3,4}$, J. Berakdar$^1$,
V. I. Ivanov$^2$, W. Dobrowolski$^5$, and V. F. Mitin$^6$}
\affiliation{ $^1$Max-Planck-Institut f\"ur Mikrostrukturphysik,
Weinberg 2, 06120 Halle, Germany\\
$^2$Institute for Problems of Materials Science, NASU,
Vilde 5, 58001 Chernovtsy, Ukraine\\
$^3$Department of Physics, A. Mickiewicz University,
Umultowska~85, 61-614~Pozna\'n, Poland\\
and Institute of Molecular Physics, PAS,
M.~Smoluchowskiego~17, 60-179~Pozna\'n, Poland\\
$^5$Institute of Physics, PAS, Al. Lotnik\'ow 32/46, 02-668 Warsaw, Poland\\
$^6$V.~Lashkaryov Institute of Semiconductor Physics, NASU,
Pr.~Nauki~41, 03028 Kiev, Ukraine}

\date{\today }

\begin{abstract}
We investigate theoretically the influence of the spin-orbit
interaction of Rashba type on the magnetoresistance of a
semiconducting ferromagnetic nanostructure with a laterally
constrained domain wall. The domain wall is assumed sharp (on the
scale of the Fermi wave length of the charge carriers). It is
shown that the magnetoresistance in such a case can be
considerably large, which is in a qualitative agreement with
recent experimental observations. It is also shown that spin-orbit
interaction may result in an increase of  the magnetoresistance.
The role of localization corrections is also briefly discussed.
\end{abstract}


\maketitle

\section{Introduction}

Rapid progress in fabrication and measurement techniques of
artificially sanitized ferromagnetic nanostructures revealed a
variety of new phenomena. For instance, in contrast to the bulk
case, it has been found that the magnetoresistance associated with
nanosize DWs can  be very
large.\cite{gregg96,garcia99,ebels00,danneau02,chopra02,ruster03}
A notable example are the experiments on Ni microjunctions, which
show that constrained DW formed at the contact of ferromagnetic
wires results in a large electrical resistance, leading thus to a
huge negative magnetoresistance.\cite{chopra02} Further insight is
provided by recent measurements of the magnetoresistance (found to
be $ \sim 2000\%$)  in semiconducting magnetic
nanoconstrictions.\cite{ruster03} This latter example is
particulary interesting insofar as the extent of  DWs  (i.e. the
width $L$ ) formed in magnetic nanoconstrictions can be on the
atomic scale\cite{bruno99} and considerably smaller than the Fermi
wavelength of charge carriers. This situation may have important
consequence as far as the the influence of DW on the transport
properties  is concerned.

On the other hand, theoretical descriptions of the transport
properties of DWs are mainly restricted to smooth DWs, typical for
bulk  or thin  film ferromagnetic
materials.\cite{levy97,gorkom99,cabrera74,brataas99,dugaev02}
Results of these studies indicate that electron scattering from
smooth DWs is rather weak, and the spin of an electron propagating
across the wall follows magnetization direction almost
adiabatically. The contribution of smooth  DWs  to electrical
resistance can be then calculated within the semiclassical
approximation, and has been found to be either positive or
negative -- but in general it is rather small. We recall, however,
that the condition for the applicability  of the semiclassical
approximation is $k_{F\uparrow (\downarrow )}L\gg 1$, where
$k_{F\uparrow }$ and $k_{F\downarrow}$ are the Fermi wavevectors
for the majority and minority electrons, respectively. This
condition is fulfilled in bulk ferromagnets.

In contrast, for $k_{F\uparrow (\downarrow )}L\leq 1$, the
semiclassical approximation is no longer valid and the scattering
of electrons from the (sharp) DWs has to be considered strong.
Therefore, various attempts have been put forward to understand
the influence of sharp DWs on transport properties. For instance,
Tagirov {\it et al}\cite{tagirov}  considered DWs in magnetic
junctions as a potential barriers independent of the electron spin
orientation. They concluded that the presence of DW results in a
large magnetoresistance. Furthermore, ballistic electron transport
through DWs was investigated
numerically.\cite{hoof99,kudrnovsky00,kudrnovsky01,yavorsky02}
Recently, the ballistic motion through a nanocontact has been
studied by Zhuravlev et al,\cite{zhuravlev03} who found a large
magnetoresistance effect due to the presence of a nonmagnetic
region within the constriction considered as a one-channel wire.

The one-dimensional model of a sharp DW has been considered in
Ref.~[\onlinecite{dugaev03}] in the limit of $k_{F\uparrow
(\downarrow )}L\ll 1$. It has been shown there that the problem
can be viewed as transmission through a spin-dependent barrier.
This results in substantial magnetoresistance that increases when
the spin polarization of electrons is enhanced. The largest
magnetoresistance is thus expected for a fully spin-polarized
electron gas. \cite{flatte01}

A question which is still not yet addressed concerns the role of
spin-orbit interaction in the scattering from a sharp DW. An
analysis of this aspect is highly desirable in view of the
relevance of spin-orbit interaction in spintronic devices, as
evidenced by recent measurements.\cite{gould} Generally, the
spin-orbit coupling can mix the spin channels, in addition to the
mixing caused by the spin-dependent scattering from the DW. As
demonstrated in this work, the presence of the spin-orbit
interaction (of the the Rashba type) results in an increase of the
magnetoresistance due to DW. In the present work we also address
briefly the role of localization corrections.

\section{Model and scattering states}

We consider a ferromagnetic narrow channel with a single magnetic
DW. In the continuous model the spin density (magnetization) is a
function of the coordinate $z$ (along the channel), ${\bf
M}(z)=\left[ M_0\sin \varphi (z),\, 0,\, M_0\cos \varphi
(z)\right]$, where $\varphi (z)$ varies continuously from zero to
$\pi $ for $z$ changing from $z=-\infty $ to $z=+\infty $.
Accordingly, the magnetization is oriented along the axis $z$ for
$z\ll -L$, and points in the opposite direction for $z\gg L$. In
what follows we assume that the DW width $L$ is less than the
Fermi wave length  $\lambda_F$ of the charge carriers. This
limiting case  is appropriate for DWs formed at constrained
magnetic contacts, in particular for low-density magnetic
semiconductors, where $\lambda_F$  can be quite large. For the
description of the conduction electrons in the semiconductor we
assume a parabolic band model. Magnetic polarization of the wire
is associated with splitting of the spin-up and spin-down electron
bands (we take the quantization axis along $z$).

Due to the spatial variation of the magnetization ${\bf M}({\bf
r})$, spin-flip scattering of electrons may occur within the
domain wall. In addition, for a sharp DW the spin-up electrons
propagating along the axis $z$ are reflected from the effective
potential barrier at $z=0$. Hence, the strongest effect of DWs on
the electronic transport can be expected in the case of a full
spin polarization of the  electron gas, i.e. when there are no
spin-down electrons at $z<0$, and no spin-up electrons at $z>0$.
This limit is reached  when $JM_0>E_F$, where $J$ is the exchange
integral, and $E_F$ is the Fermi energy in the absence of
magnetization. We recall that $E_F$ characterizes the total
electron density $n$ of the semiconducting material,
$n=(2mE_F)^{3/2}/3\pi ^2\hbar ^3$, where $m$ is the electron
effective mass. Hence, the condition ($JM_0>E_F$) of full spin
polarization becomes particularly satisfied when  a depletion
region near the DW exists.

As mentioned above, the condition of sharp DW means that the wall
width is smaller than the electron Fermi wavelength, i.e.
$k_FL<1$, where $k_F$ is the electron Fermi wavevector. This
condition can be easily fulfilled in semiconductors, especially in
the case of low electron concentration. In addition, when DW is
laterally constrained, the number of quantum transport channels
can be reduced substantially. In the extreme case only a single
conduction channel can be active. The corresponding condition is
$k_FL_c<1$, where $L_c$ is the wire width. This condition can be
easily obeyed in semiconductors with low density of carriers.

An important element of the model is the presence of spin-orbit
interaction. Under the condition of full spin polarization, the
spin-flip scattering provides mixing of different spin channels,
that is responsible for the transfer of electrons through the
domain wall. Thus, one can expect strong influence of spin-orbit
interaction on the total resistance. In the following we assume
the spin-orbit interaction in the form of Rashba term. Such an
interaction is usually associated with the asymmetric form of the
confining potential leading to size quantization in quantum wells
and wires. The model Hamiltonian we analyze in this work has the
form
\begin{equation}
\label{1}
H=-\frac{\hbar ^2}{2m}\, \frac{d^2}{dz^2}-JM_z(z)\,
\sigma _z-J M_x(z)\, \sigma _x\, +i\alpha \, \sigma _x\,
\frac{d}{dz}\; ,
\end{equation}
where $\alpha $ is the parameter of spin-orbit interaction,
whereas $\sigma _x$ and $\sigma _z$ are the Pauli matrices. We
choose the axis $x$ to be normal to the wire and assume that the
magnetization in the wall rotates in the $x$-$z$ plane. The Rashba
spin-orbit interaction in Eq.~(1) corresponds to the axis $y$
perpendicular to the substrate plane. The magnetization vector
rotates then in the substrate plane. Although the one-dimensional
model describes only a single-channel quantum wire, it is
sufficient to account qualitatively for some of the recent
observations. In addition, the present model can be  generalized
straightforwardly to the case of a wire with more conduction
channels (large width and/or higher carrier concentration).

Our treatment is based on the scattering states. For electrons
incident from left to right, the asymptotic form of such states
(taken sufficiently far from DW, $\left| z\right| \gg L$) is
\begin{eqnarray}
\label{2}
\psi _{kR}(z)=\frac{e^{ikz}}{D_k} \left(
\begin{array}{c}
M_k \\
\alpha k
\end{array} \right)
+\frac{r\, e^{-ikz}}{D_k} \left( \begin{array}{c}
M_k \\
-\alpha k
\end{array} \right)
\nonumber \\
+\frac{r_f\, e^{\kappa z}}{D_\kappa }\;
\left( \begin{array}{c}
i\alpha \kappa \\
M_\kappa
\end{array} \right) ,\hskip0.3cm z\ll -L,
\end{eqnarray}
\begin{eqnarray}
\label{3}
\psi _{kR}(z)
=\frac{t_f\, e^{ikz}}{D_k} \left(
\begin{array}{c}
\alpha k \\
M_k
\end{array} \right)
+\frac{t\, e^{-\kappa z}}{D_\kappa }\;
\left( \begin{array}{c}
M_\kappa \\
-i\alpha \kappa
\end{array} \right) ,
\nonumber \\
\hskip1cm  z\gg L.
\end{eqnarray}
In Eqs.~(2) and (3) $k$ and $\kappa$ are defined as $k=\left[
2m(E+M)\right] ^{1/2}/\hbar $ and $\, \kappa =\left[
2m(M-E)\right] ^{1/2}/\hbar $, respectively, whereas the other
parameters are $\, M_k=M+\left( M^2+\alpha ^2k^2\right) ^{1/2}$,
$M_\kappa =M+\left( M^2-\alpha ^2\kappa ^2\right) ^{1/2}$,
$D_k=\left( M_k^2+\alpha ^2k^2\right) ^{1/2}$, and $D_\kappa
=\left( M_\kappa ^2+\alpha ^2\kappa ^2\right) ^{1/2}$. Here, $M$
is defined as $M=JM_0$ and $E$ denotes the electron energy.

Due to spin-orbit interaction, electron states are superpositions
of spin-up and spin-down components. For simplicity, we call them
in the following either spin-up or spin-down waves, because they
reduce to such waves in the limit of vanishing spin orbit
interaction. Thus, the scattering state (2),(3) describes the
spin-up wave incident from $z=-\infty $ to the right, which is
partially reflected and partially transmitted into the spin-up and
spin-down channels. The coefficients $t$ and $t_f$ are the
transmission amplitudes without and with spin reversal,
respectively, whereas $r$ and $r_f$ are the corresponding
reflection amplitudes. Even though there are no minority carriers
far from the domain wall, the corresponding wavefunction
components exist in the vicinity of the domain wall and decay
exponentially in the bulk. Similar form applies to  the scattering
states $\psi _{kL}$ describing electrons incident from the right
to the left.

When $kL\ll 1$, the reflection and transmission coefficients can
be calculated analytically. Upon integrating the Schr\"odinger
equation $H\psi =E\psi $ (with the Hamiltonian given by Eq.~(1))
from $z=-\,\delta $ to $z=+\,\delta $, and assuming $L\ll \delta
\ll k^{-1}$, one obtains
\begin{equation}
\label{4} \left. \frac{d\psi _{kj}}{dz}\right| _{z=+\delta }
-\left. \frac{d\psi _{kj}}{dz}\right| _{z=-\delta }
+\frac{2m\lambda }{\hbar }\; \sigma _x \, \psi _{kj}(z=0)=0
\end{equation}
for each scattering state ($j=R,L$), where
\begin{equation}
\label{5}
\lambda\simeq \frac{J}{\hbar }
\int _{-\infty }^\infty dz\; M_x(z).
\end{equation}
Equation (4) has the form of a spin-dependent condition for
electron transmission through a $\delta $-like potential barrier
located at $z=0$ and was obtained assuming $kL\ll 1$. The magnitude of
the parameter $\lambda$ defined in Eq.~(5) can be estimated as
$\lambda\simeq JM_0L/\hbar = ML/\hbar $.

Using the full set of scattering states, together with the wave
function continuity condition, one can find a set of equations for
the transmission amplitudes $t$ and $t_f$. Since the wavefunction
component corresponding to conserved electron spin decays
exponentially away from the wall, only the spin-flip amplitude
$t_f$ determines the electric current in the wire. Let us denote
the velocity of  the incident electrons  by $v$, $v=k/m$, and by
$\nu$ the corresponding quantity for the exponentially decaying
wave component, $\nu =\kappa /m$. From the Schr\"odinger equation
two equations are deduced for the transmission amplitudes $t$ and
$t_f$, namely

\begin{eqnarray}
\label{6}
\left[
ivM_k-\nu M_\kappa
-2i\lambda \alpha \kappa
-\frac{\alpha ^2\kappa (v+i\nu )(M_\kappa k-iM_k\kappa )}
{i\alpha ^2k\kappa +M_kM_\kappa }\right]
\nonumber \\
\times \frac{t}{D_\kappa }
+\left[
2i\alpha v k+2\lambda M_k
+\frac{\alpha \kappa (v-i\nu )(\alpha ^2k^2+M_k^2)}
{i\alpha ^2k\kappa +M_kM_\kappa }\right]
\frac{t_f}{D_k}
\nonumber \\
=\frac{2ivM_k}{D_k}
-\frac{2i\alpha ^2k\kappa (D_k+M_k)}
{D_k(i\alpha ^2k\kappa +M_kM_\kappa )}\; ,\hskip0.3cm
\end{eqnarray}

\begin{eqnarray}
\label{7}
\left[
\frac{i\alpha vkM_\kappa}{M_k}
+\frac{\alpha \nu (\alpha ^2k^2+M_kM_\kappa )(M_\kappa k-iM_k\kappa )}
{M_k(i\alpha ^2k\kappa +M_kM_\kappa )}\hskip0.5cm
\right. \nonumber \\ \left.
-i\alpha \nu \kappa
-2\lambda M_\kappa
\right]
\frac{t}{D_\kappa }
+\left[
\frac{iv\alpha ^2k^2}{M_k}
-ivM_k-2\alpha \lambda k \hskip0.5cm
\right. \nonumber \\ \left.
+\frac{\nu(\alpha ^2k^2+M_k^2)^2}
{M_k(i\alpha ^2k\kappa +M_kM_\kappa )}\right]
\frac{t_f}{D_k}
=\frac{2\alpha k\nu (\alpha ^2k^2+M_kM_\kappa )}
{D_k(i\alpha ^2k\kappa +M_kM_\kappa )}\; .
\hskip0.3cm
\end{eqnarray}

In the absence of spin-orbit interaction, $\alpha =0$, one finds
\begin{equation}
\label{8}
t=\frac{2v(v+i\nu )}{\left( v+i\nu \right) ^2+4\lambda ^2},\hskip0.3cm
t_f=\frac{4i\lambda v}{\left( v+i\nu \right) ^2+4\lambda ^2}.
\end{equation}
In the limit of $\nu \gg v$ and $\lambda \ll \nu $ (low density of
carriers and small spin-orbit interaction) another
limiting formula is derived
\begin{equation}
\label{9}
t_f=-\frac{4iv\lambda ^2}{\nu ^2\left( \lambda
-i\alpha \nu \kappa /M\right) }\; .
\end{equation}
In  general, the coefficient $t_f$ can be found
analytically but the corresponding formula is rather cumbersome.

In the limit of $\lambda \rightarrow 0$ (very thin DW), the
transmission through the wall vanishes, which corresponds to the
complete reflection of electrons from the wall. Thus, at first
glance one might expect that a nonzero spin-orbit interaction
mixes the spin channels and leads to nonvanishing transmission
through the wall, even in the limit of very thin domain wall. This
is however not the case since the matching condition for the wave
functions at $z<L$ and $z>L$ requires that both incident and
transmitted waves are certain superpositions of spin-up and
spin-down components. On the other hand, equation (9) indicates
that transmission through the wall decreases with increasing
strength of the spin-orbit interaction.

\section{Resistance of the domain wall}

To calculate the conductance of the system, we use the
B\"uttiker-Landauer formula, which can be   simplified substantially
due to the suppression of all channels, but spin-flip through the wall.
(The derivation of such a formula for transmission through the
wall in the case of all nonvanishing channels has been done in
Ref. ~[\onlinecite{dugaev03}].) Thus, one obtains
\begin{equation}
\label{10}
G=\frac{e^2}{2\pi \hbar }\; \left| t_f\right| ^2.
\end{equation}
Due to the asymptotic current conservation, the conductivity is
determined by the propagating (non-decaying) component of the
transmitted wave. Using Eq.~(8) one finds for vanishing spin-orbit
interaction
\begin{equation}
\label{11}
G =\frac{8e^2}{\pi \hbar}\; \frac{\lambda ^2\, v^2}
{\left( v^2-\nu^2+4\lambda ^2\right) ^2+4v^2\, \nu ^2}.
\end{equation}
Here,  all the velocities are taken at the Fermi level.

\begin{figure}
\includegraphics[scale=0.5]{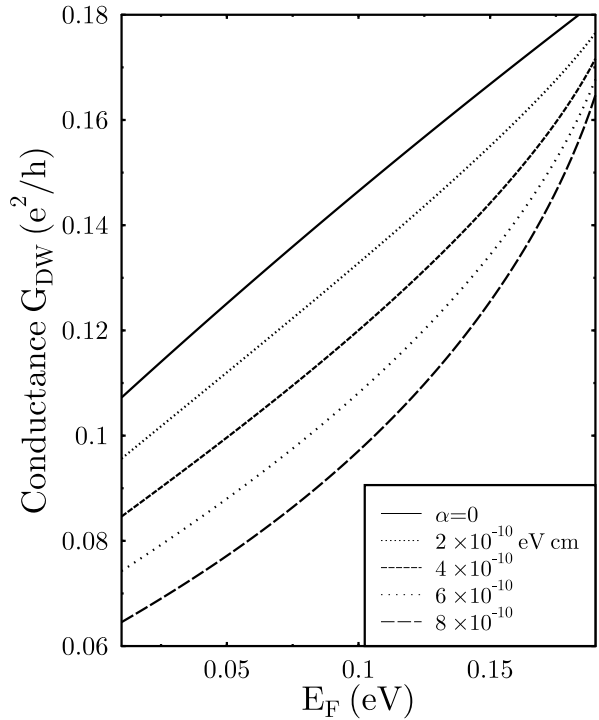}
\caption{Conductance of a magnetic wire with a single domain wall
vs. Fermi energy of electrons. Different curves correspond to
different values of the spin-orbit coupling parameter $\alpha $.}
\end{figure}

\begin{figure}
\includegraphics[scale=0.5]{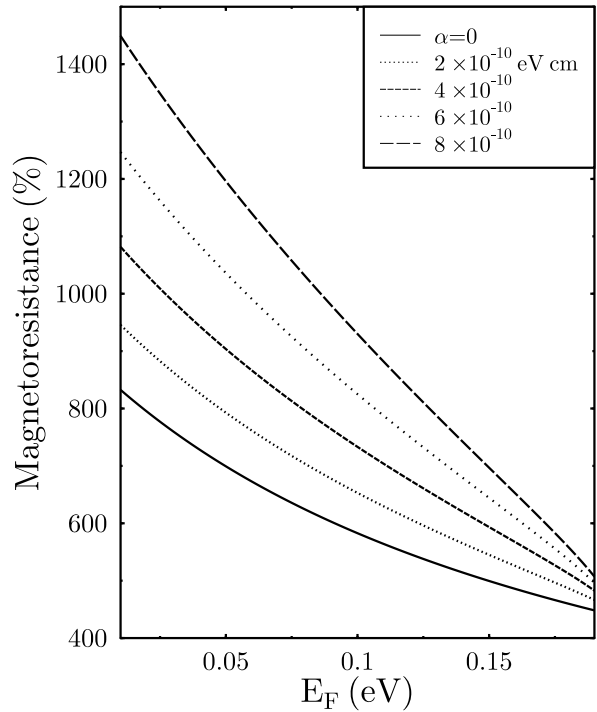}
\caption{Magnetoresistance of the wire with a domain wall vs.
Fermi energy for different values of $\alpha $.}
\end{figure}

Figure 1 shows  the calculated dependence of the electrical
conductance on the Fermi-energy $E_F$ in the general case. The
calculations were performed  assuming the following values of the
relevant parameters: $m=0.6\, m_0$ (where $m_0$ is the free
electron mass), $JM_0=0.2$~eV, and $L=10^{-8}$~cm. These
parameters correspond to GaMnAs semiconductor, and satisfy the
condition $JM_0>E_F$ for $E_F<0.2$~eV.

We can estimate the magnitude of parameter $\alpha $ by taking the
value of the spin-orbit (SO) splitting $\Delta E_{SO}\simeq \alpha
k$, where the momentum $k$ is related to the density of carriers
$N_s$ as $k=(2\pi N_s)^{1/2}$. Assuming $\Delta E_{SO}=0.5$~meV
for $N_s=10^{11}$~cm$^{-2}$ as a characteristic value for
GaAs-GaAlAs heterostructures,\cite{pfeffer95} one obtains $\alpha
\simeq 6.3\times 10^{-10}$~eV$\cdot $cm.

From Fig.~1 it is clear that the conductance increases
monotonically with increasing $E_F$ because the barrier is felt
smaller by electrons having higher energy. Furthermore, the
conductance of a magnetic wire with DW diminishes with increasing
strength of the spin-orbit interaction.

The dependence of the magnetoresistance on the Fermi energy $E_F$ is
presented in Fig.~2 for different values of the parameter $\alpha
$. The magnetoresistance is calculated with respect to the state
without DW, $MR=R_{DW}/R_0-1$, where $R_{DW}$ is the resistance of
the wire with DW and $R_0=2\pi \hbar /e^2$ is its resistance in
the absence of the wall (only spin-up channel is active). For our
choice of parameters, the magnetoresistance is rather high and
increases substantially with spin-orbit interaction.

The magnetoresistance measurements on magnetic semiconductors are
usually performed at low temperatures because the corresponding
Curie temperature is rather low. At such conditions, one can
expect a significant contribution of the localization corrections
to the conductivity. The role of the localization in the case of
smooth DWs (for $kL\gg 1$) has been studied before,
\cite{tatara97,lyanda98} and it was shown that the localization
corrections are suppressed by an effective gauge field of the
wall. This means that the contribution of the wall to resistance
is negative, and the corresponding magnetoresistance is positive.

We have analyzed the role of localization corrections in the case
of sharp DW. Qualitively, it can be described as the DW induced
suppression of the quantum interference in triplet Cooperon
channel.\cite{lee85} The singlet channel in ferromagnets is
strongly suppressed by the internal magnetization.\cite{dugaev01}
The suppression of the interference by DWs is related to dephasing
of the wave function of electron transmitted through the
barrier.\cite{jonkers99,tatara01} If the transmission through the
wall is small, the corresponding dephasing length roughly equals
to the distance of electron moving from a point $z$ (within the
constriction) to the domain wall position ($z=0$), and the
dephasing time is $\tau _{dw}(z)\sim z^2/D$, where $D$ is the
diffusion coefficient. After averaging over $z$ of the
localization correction $\delta G(z)$, we find that the
characteristic dephasing length $L_0$ is the constriction length
itself, $\delta G_{dw}\simeq -e^2L_0/\pi \hbar $. In the case of
sharp DWs, the localization correction diminishes the
magnetoresistance due to the reflection from the wall, since it
has a different sign.

\section{Conclusions}

We have presented a theoretical description of the resistance of a
semiconducting magnetic nanojunction with a constrained DW in the
case of a full spin polarization of electron gas. In the limit of
$kL\ll 1$, the electron transport across the wall was treated
effectively as electron tunneling through a spin-dependent
potential barrier. For such a narrow and constrained DW, the
electron spin does not follow adiabatically the magnetization
direction, but its orientation is rather fixed. However, DW
produces some mixing of the spin channels. The spin-orbit
interaction essentially enhances the magnetoresistance, whereas
the localization corrections play the opposite role. However, the
localization corrections can be totally suppressed by the
spin-orbit interaction.\cite{dugaev01} This indicates that the
spin-orbit interaction can play an important role and can lead to
large enhancement of the magnetoresistance effect.

\begin{acknowledgments}
This work is supported by Polish State Committee for Scientific
Research under Grants Nos.~PBZ/KBN/044/P03/2001 and 2~P03B~053~25,
and also by INTAS Grant No.~00-0476.
\end{acknowledgments}


\begin{thebibliography}{99}

\bibitem[\dag ]{email}
Email address: vdugaev@mpi-halle.de

\vskip0.3cm
\bibitem{kent01}
A. D. Kent, J. Yu, U. R\"udiger, and S. S. P. Parkin, J. Phys.
Cond. Matter {\bf 13}, R461 (2001).

\bibitem{hong98}
K. Hong and N. Giordano, J. Phys. Cond. Matter {\bf 13}, L401
(1998).

\bibitem{rudiger98}
U. Ruediger, J. Yu, S. Zhang, A. D. Kent, and S. S. P. Parkin,
Phys. Rev. Lett. {\bf 80}, 5639 (1998).

\bibitem{kent99}
A. D. Kent, U. R\"udiger, J. Yu, L. Thomas, and S. S. P. Parkin,
J. Appl. Phys. {\bf 85}, 5243 (1999).

\bibitem{gregg96}
J. F. Gregg, W. Allen, K. Ounadjela, M. Viret, M. Hehn,
S. M. Thompson, and J. M. D.~Coey, Phys. Rev. Lett. {\bf 77}, 1580
(1996).

\bibitem{garcia99}
N. Garcia, M. Mu$\tilde{\rm n}$oz, and Y. W. Zhao, Phys. Rev.
Lett. {\bf 82}, 2923 (1999).

\bibitem{ebels00}
U. Ebels, A. Radulescu, Y. Henry, L. Piraux, and K. Ounadjela,
Phys. Rev. Lett. {\bf 84}, 983 (2000).

\bibitem{danneau02}
R.~Danneau, P.~Warin, J.P. Attan\'e, I. Petej, C. Beign\'e,
C. Fermon, O. Klein, A. Marty, F. Ott, Y. Samson, and M. Viret,
Phys. Rev. Lett. {\bf 88}, 157201 (2002).

\bibitem{chopra02}
H.~D.~Chopra and S.~Z.~Hua, Phys. Rev. B {\bf 66}, 020403(R)
(2002).

\bibitem{ruster03}
C. R\"uster, T. Borzenko, C. Gould, G. Schmidt, L. W. Molenkamp,
X. Liu, T. J. Wojtowicz, J. K. Furdyna, Z. G. Yu, and M. E.
Flatt\'e, Phys. Rev. Lett. {\bf 91}, 216602 (2003).

\bibitem{bruno99}
P.~Bruno, Phys. Rev. Lett. {\bf 83}, 2425 (1999).

\bibitem{levy97}
P. M. Levy and S. Zhang, Phys. Rev. Lett. {\bf 79}, 5110 (1997).

\bibitem{gorkom99}
R. P. van Gorkom, A. Brataas, and G.~E.~W. Bauer,
Phys. Rev. Lett. {\bf 83}, 4401 (1999).

\bibitem{cabrera74}
G.~G.~Cabrera and L.~M.~Falicov,
Phys. Status Solidi B {\bf 61}, 539 (1974);
{\bf 62}, 217 (1974).

\bibitem{brataas99}
A.~Brataas, G.~Tatara, and G. E. W. Bauer,
Phys. Rev. B {\bf 60}, 3406 (1999).

\bibitem{dugaev02}
V.~K.~Dugaev, J.~Barna\'s, A. \L usakowski, and \L . A. Turski,
Phys. Rev. B {\bf 65}, 224419 (2002).

\bibitem{tagirov}
L.~R.~Tagirov, B. P. Vodopyanov, and K. B. Efetov,
Phys. Rev. B {\bf 65}, 214419 (2002); {\bf 63}, 104428 (2001);
L. R. Tagirov, B. P. Vodopyanov, and B. M. Garipov,
J. Magn. Magn. Mater. {\bf 258-259}, 61 (2003).

\bibitem{hoof99}
J. B. A. N. van Hoof, K. M. Schep, A. Brataas, G.~E.~W.~Bauer, and
P. J. Kelly, Phys. Rev. B {\bf 59}, 138 (1999).

\bibitem{kudrnovsky00}
J. Kudrnovsky, V. Drchal, C. Blaas, P. Weinberger, I. Turek,
and P. Bruno, Phys. Rev. B {\bf 62}, 15084 (2000).

\bibitem{kudrnovsky01}
J. Kudrnovsky, V. Drchal, I. Turek, P. Streda, and P.~Bruno,
Surf. Sci. {\bf 482-485}, 1107 (2001).

\bibitem{yavorsky02}
B. Yu. Yavorsky, I. Mertig, A. Ya. Perlov, A. N. Yaresko, and
V. N. Antonov, Phys. Rev. B {\bf 66}, 174422 (2002).

\bibitem{zhuravlev03}
M. Ye. Zhuravlev, E. Y. Tsymbal, S. S. Jaswal, A. V. Vedyayev,
and B. Dieni, Appl. Phys. Lett. {\bf 83}, 3534 (2003).

\bibitem{dugaev03}
V. K. Dugaev, J. Berakdar, and J. Barna\'s,
Phys. Rev. B {\bf 68}, 104434 (2003).

\bibitem{flatte01}
M. E. Flatt\'e and G. Vignale, Appl. Phys. Lett. {\bf 78}, 1273 (2001).

\bibitem{gould}
C. Gould, C. R\"uster, T. Jungwirth, E. Girgis, G. M. Schott, R.
Giraud, K. Brunner, G. Schmidt, and L. W. Molenkamp,
cond-mat/0407735.

\bibitem{pfeffer95}
P. Pfeffer and W. Zawadzki, Phys. Reb. B {\bf 52}, R14332 (1995).

\bibitem{tatara97}
G. Tatara and H. Fukuyama, Phys. Rev. Lett. {\bf 78}, 3773 (1997).

\bibitem{lyanda98}
Y. Lyanda-Geller, I. L. Aleiner, and P. M. Goldbart, Phys. Rev.
Lett. {\bf 81}, 3215 (1998).

\bibitem{lee85}
P. A. Lee and T. V. Ramakrishnan,
Rev. Mod. Phys. {\bf 57}, 287 (1985).

\bibitem{dugaev01}
V. K. Dugaev, P. Bruno and J. Barna\'s,
Phys. Rev. B {\bf 64}, 144423 (2001).

\bibitem{jonkers99}
P.~A.~E. Jonkers, S. J. Pickering, H. De Raedt, and G. Tatara,
Phys. Rev. B {\bf 60}, 15970 (1999).

\bibitem{tatara01}
G. Tatara, Int. J. Mod. Phys. B {\bf 15}, 321 (2001).

\end{thebibliography}
\end{document}